\documentclass[]{spie}  

\usepackage{amsmath,amsfonts,amssymb}
\usepackage{siunitx}
\usepackage{graphicx}

\usepackage{caption}
\usepackage{subcaption}
\usepackage{gensymb}
\usepackage{float}

\usepackage{lineno}

\usepackage[colorlinks=true, allcolors=blue]{hyperref}

\title{First laboratory and on-sky results of an adaptive secondary mirror with TNO-style actuators on the NASA Infrared Telescope Facility}

\author[a]{Ellen Lee}
\author[a]{Mark Chun}
\author[b]{Olivier Lai}
\author[a]{Ruihan Zhang}
\author[c]{Max Baeten}
\author[c]{Arjo Bos}
\author[c]{Matias Kidron}
\author[c]{Fred Kamphues}
\author[c]{Stefan Kuiper}
\author[c]{Wouter Jonker}
\author[a]{Michael Connelley}
\author[a]{John Rayner}
\author[a]{Alan Ryan}
\author[d]{Philip Hinz}
\author[d]{Rachel Bowens-Rubin}
\author[a]{Charles Lockhart}
\author[a]{Michael Kelii}
\affil[a]{University of Hawai'i at Manoa - Institute for Astronomy, Honolulu, HI, USA}
\affil[b]{Observatoire de la Côte d'Azur, Nice, France}
\affil[c]{TNO, Delft, Netherlands}
\affil[d]{University of California - Santa Cruz, Santa Cruz, CA, USA}

\authorinfo{Further author information: (Send correspondence to E.L.)\\E.L.: E-mail: ellenlee@hawaii.edu\\M.C.: E-mail: markchun@hawaii.edu}

\begin{document} 
\maketitle

\begin{abstract}
We are developing an adaptive secondary mirror (ASM) that uses a new actuator technology created by the Netherlands Organization for Applied Scientific Research (TNO). The TNO hybrid variable reluctance actuators have more than an order of magnitude better efficiency over the traditional voice coil actuators that have been used on existing ASMs and show potential for improving the long-term robustness and reliability of ASMs. To demonstrate the performance, operations, and serviceability of TNO’s actuators in an observatory, we have developed a 36-actuator prototype ASM for the NASA Infrared Telescope Facility (IRTF) called IRTF-ASM-1. IRTF-ASM-1 provides the first on-sky demonstration of this approach and will help us evaluate the long-term performance and use of this technology in an astronomical facility environment. We present calibration and performance results with the ASM in a Meniscus Hindle Sphere lens setup as well as preliminary on-sky results on IRTF. IRTF-ASM-1 achieved stable closed-loop performance on-sky with H-band Strehl ratios of 35-40\% in long-exposure images under a variety of seeing conditions.  
 
\end{abstract}


\section{Introduction}

Adaptive secondary mirrors can provide a number of advantages that make them highly attractive to observatories. Because an ASM is incorporated into the telescope itself, it can provide drastically improved image quality to every instrument without sacrificing throughput and thermal emissivity. Unfortunately, ASMs are currently uncommon due to the complexity of their design which is primarily driven by the available actuator technology.

TNO has developed hybrid variable reluctance (HVR) actuators with an 80-fold increase in efficiency over the traditional voice coil actuators that have been used on existing ASMs \cite{2020SPIE11448E..5VB}. The increased efficiency allows for a thicker facesheet which greatly increases the robustness of the ASM. The actuators themselves are also more robust and provide a linear force output over a large dynamic range. This linearity means that the HVR actuators do not require position feedback from capacitive sensors resulting in additional power and complexity reductions.  The combination of a simple drive circuitry and the high efficiency of the HVR approach dramatically reduces the power dissipation and eliminates the need for a complex (active) cooling system. It also potentially allows the actuators to be packed more closely together.

The development of an ASM with TNO's HVR actuators has been ongoing for several years. The actuators have been demonstrated in the lab on large, flat deformable mirrors including DM3 and FLASH \cite{2020SPIE11448E..5VB, 2021SPIE11823E..1RB, bowens2023performance}. Originally, the first on-sky demonstration of these actuators was planned for the University of Hawaii 2.2-meter telescope (UH2.2) \cite{kuiper2019adaptive, 2020SPIE11448E..1EC, 2022SPIE12185E..7UC}. As of writing, UH2.2's ASM is being integrated at TNO and is scheduled for delivery near the end of 2024.  Furthermore, TNO is currently developing an improved actuator design that will enable reduced inter-actuator spacing. These actuators are also being investigated for other observatories including the Automated Planet Finder Telescope and the W. M. Keck Observatory \cite{hinz2020developing, bowens2022adaptive}.

\begin{figure}
    \centering
    \includegraphics[scale=0.2]{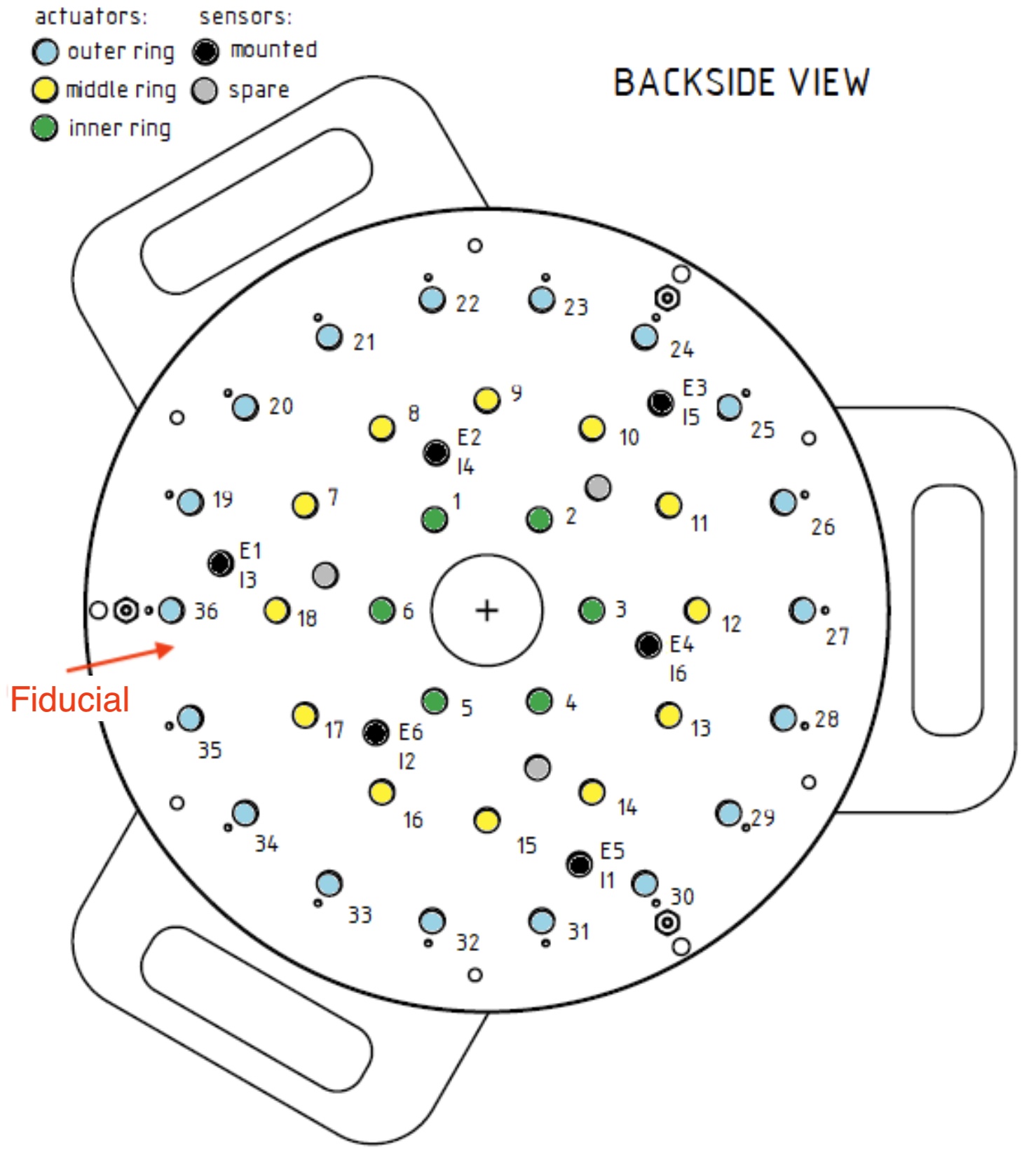}
    \caption{Layout of IRTF-ASM-1. There are 36 actuators distributed into three rings (labeled 1-36). Six capacitive sensors are mounted in the assembly (designated by I1-I6) to monitor the distance between the body of the ASM and the back of the optical facesheet. These are used for functionality tests and are not part of the adaptive optics closed-loop control of the actuators. We use a fiducial to help identify the orientation of the ASM. Refer to \citenum{2024SPIE_arjo} for further information on the design of IRTF-ASM-1.}
    \label{fig:irtf-1}
\end{figure}

The NASA IRTF-ASM-1 was conceived to bring the technology to an astronomical telescope as quickly as possible. It leverages the design and parts for the UH2.2 adaptive secondary mirror project and TNO prototype deformable mirrors to enable a 36-actuator ASM that can provide good wavefront correction in the near-infrared on the NASA IRTF 3.2-meter telescope on Maunakea, Hawaii. The infrared-optimized NASA IRTF has a slow f/38 beam with a relatively small (\qty{244}{\milli\meter}) diameter secondary mirror. This simplified the production of the optical facesheet compared to the UH2.2 ASM's \qty{620}{\milli\meter} diameter secondary.   A layout of the IRTF-ASM-1's 36 actuators is shown in Fig.~\ref{fig:irtf-1}. The project took just one year from start of funding to its delivery to Hawaii.  Lab testing and on-sky testing were completed in two months with a successful on-sky demonstration in April this year. Table ~\ref{tab:timeline} shows a timeline of the project.

\begin{table}[]
    \centering
    \begin{tabular}{|r|l|}
        \hline
         Jul 2022 & Initial concept of IRTF-ASM-1 discussed at SPIE. \\
         \hline
         Q1 2023 & Funding secured (TNO, IRTF, NSF) in January. Begin project. \\
         \hline
         Q2 2023 & Design of the ASM complete, begin fabrication. \\
         \hline
         Q3 2023 & Integration of IRTF-ASM-1 at TNO. AO subsystem integration with DM3/FLASH.\\
         \hline
         Q4 2023 & TNO lab and acceptance testing. Lab AO setup integration in Hawai'i.\\
         \hline
         Jan 2024 & Lab setup preparations in Hilo using a static IRTF secondary mirror. \\
         \hline
         Feb 2024 & Delivery of IRTF-ASM-1 to Hilo. Begin lab testing. \\
         \hline
         Apr 2024 & IRTF-ASM-1 on-sky. Closed loop on alf Boo during the first night. \\
         \hline
    \end{tabular}
    \caption{The highly compressed timeline of IRTF-ASM-1.}
    \label{tab:timeline}
\end{table}

This manuscript is one of two papers about IRTF-ASM-1 and TNO's HVR actuators that is being submitted to this conference; refer to \citenum{2024SPIE_suz} for information about the dynamic response of the actuators and \citenum{2024SPIE_arjo} for the design and integration of IRTF-ASM-1. Also refer to \citenum{2024SPIE_phil} for information about slumping the facesheet. In this paper, we focus on static lab tests of IRTF-ASM-1 and its on-sky performance. We begin by describing the optical design of the test bench in Sec.~\ref{sec:design}. Then, in Sec.~\ref{sec:lab_tests} we discuss the static performance of IRTF-ASM-1 measured in the lab, and finally, we present the first on-sky, closed-loop results of the ASM in Sec.~\ref{sec:sky}.

\section{Optical bench design} \label{sec:design}
\begin{figure}[H]
    \centering
    \includegraphics[width=\linewidth]{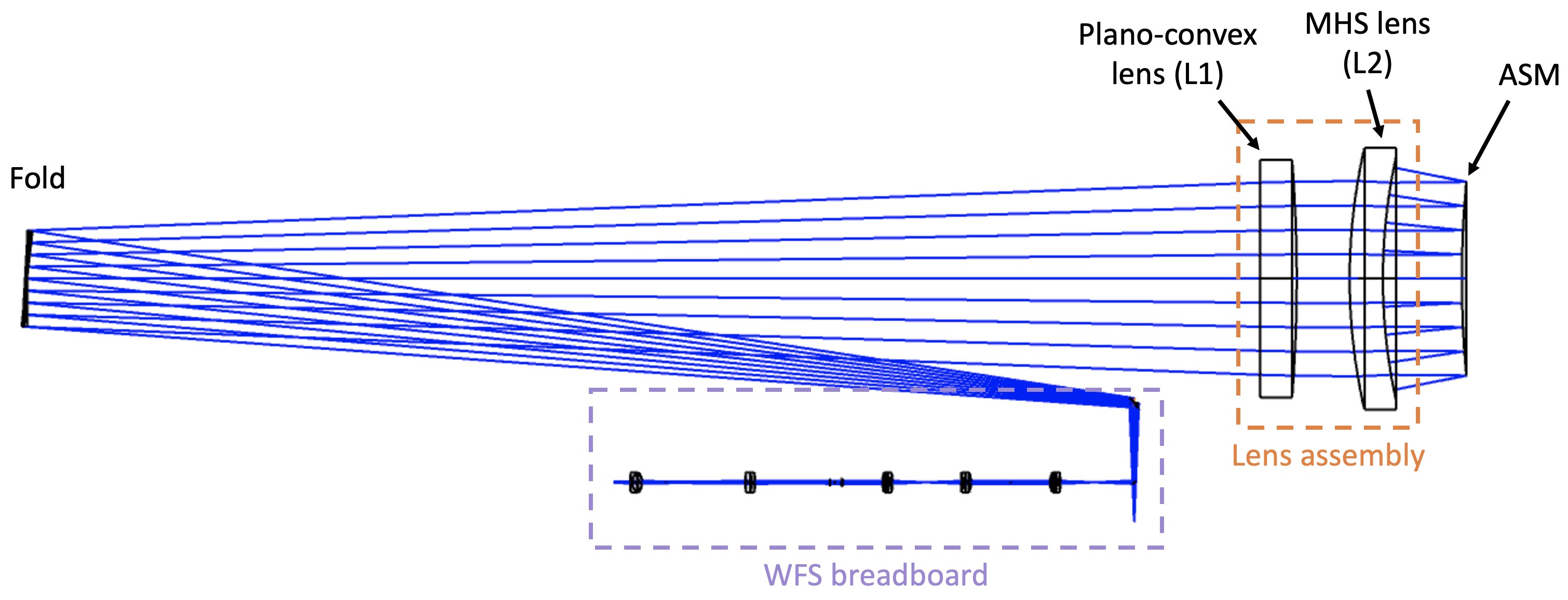}
    \caption{A ray trace of the test bench for IRTF-ASM-1. Orange: The lens assembly. A plano-convex lens (L1) is used to reduce the f-ratio of the beam. The concave side of the MHS lens (L2) has a 33\% reflective coating and serves as the Hindle surface in the traditional Hindle sphere test. Purple: The wavefront sensor (WFS) breadboard. This raytrace only shows the low-order $12\times12$ Shack-Hartmann WFS that is used to drive the ASM.}
    \label{fig:bench_zemax}
\end{figure}

Inspired by Ref.~\citenum{2008SPIE.7015E..65S} and Ref.~\citenum{1974ESOTR...3.....W}, we used a Meniscus Hindle Sphere (MHS) lens as the basis for the test setup (adopting the terminology from Ref.~\citenum{1974ESOTR...3.....W}). A Hindle test involves placing a spherical mirror such that its center of curvature overlaps with the virtual focus of the convex, hyperbolic secondary mirror, which creates a return beam that follows the same path as the test beam. The MHS setup improves upon the traditional Hindle sphere test by replacing the spherical mirror with a meniscus lens that has a partially reflective coating on its concave face. This drastically decreases the size of the Hindle sphere. Figure~\ref{fig:bench_zemax} shows a ray trace of the test bench.

The main components of the MHS setup are as follows:

\begin{itemize}
    \item ASM mount: The mount holds IRTF-ASM-1 upright and provides a safe way to handle and align the ASM.
    \item Lens assembly: This holds the two lenses L1 and L2. L2 serves as the MHS lens. L1 increases the F-ratio of the beam from f/12.7 to f/38.3 at the ASM, which matches the telescope; L1 prevents the test setup from becoming prohibitively large. Although the assembly is built to allow one to change the centering of the lenses, the lenses are not meant to move with respect to each other once they are in place.
    \item Wavefront sensor (WFS) breadboard: This package is eventually mounted to the telescope and contains all other optical components needed to run the system, e.g., wavefront sensors, a pupil steering mirror, and the science camera. Further information about the WFS package is provided in \citenum{2024SPIE_suz}.
\end{itemize}

\subsection{Alignment}
\begin{figure}[H]
    \centering
    \includegraphics[width=\linewidth]{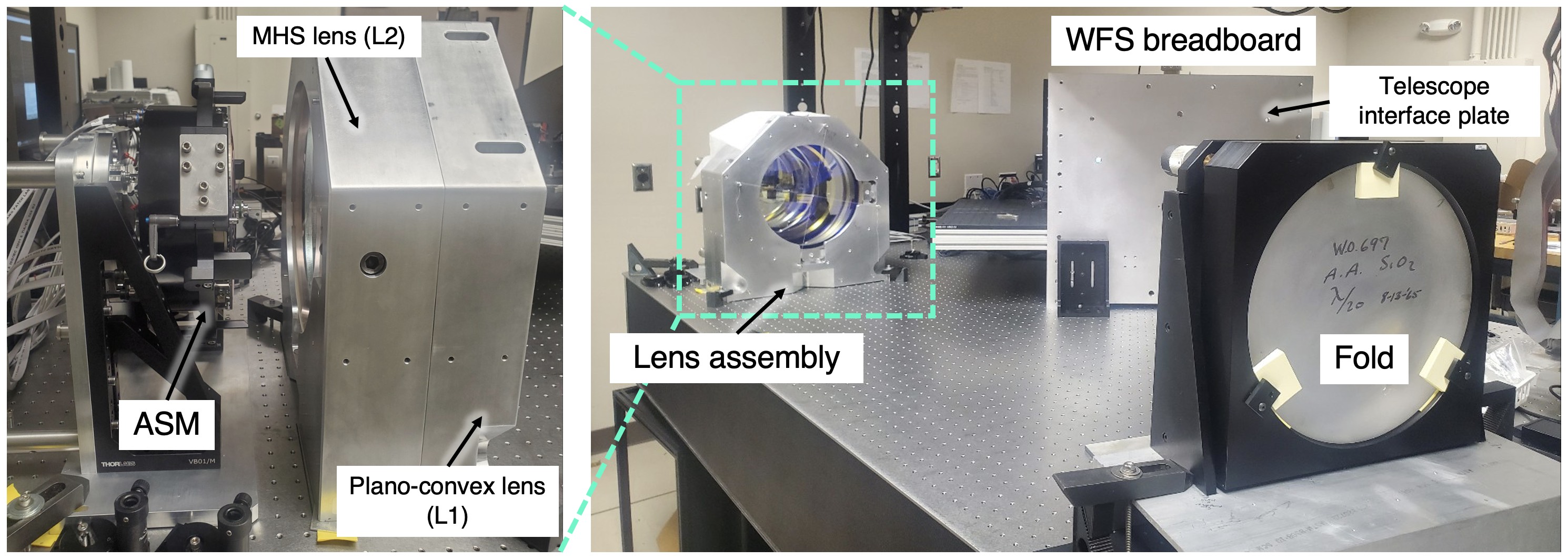}
    \caption{Pictures of the MHS test bench. The ASM sits behind the lens assembly. The mounts for L1 and L2 are bolted together to form the lens assembly. A large fold mirror sends light between the WFS breadboard and the rest of the setup. The view of the WFS breadboard is mostly obscured by the telescope interface plate, which defines the reference axis of the system.}
    \label{fig:bench_pics}
\end{figure}

We started to put together the optical bench at the beginning of January 2024. In lieu of IRTF-ASM-1, which arrived in February, we used a spare IRTF secondary mirror with a known surface figure (henceforth referred to as the "static secondary") to test the alignment procedure and exercise the WFS breadboard. Shown in Fig.~\ref{fig:bench_pics}, we chose the bore of the telescope interface plate as the reference axis of the system. This is natural because, as the name suggests, the telescope interface plate is used to mount instruments to the telescope and is where the IRTF optical system and the WFS breadboard come together.

First, we aligned the WFS breadboard by using an on-board monochromatic point source such that the WFS tube ran parallel to the breadboard and the distance between each component was correct. The configuration of the WFS breadboard is discussed further in \citenum{2024SPIE_suz}. We then decided to use an alignment telescope to bring all components of the MHS system onto the reference axis. The general procedure to align the WFS package to the lens assembly and the static secondary was as follows:

\begin{enumerate}
    \item Mount the alignment telescope to the breadboard such that it can look through the axis of the WFS tube.
    \item Bring the alignment telescope onto the axis of the WFS tube. Use a target to check the centering and a flat to view the tilt via the autocollimation rings on the alignment telescope.
    \item To avoid vignetting, check that each of the flats and the dichroic are centered on the alignment telescope.
    \item Use an alignment target and a flat to view the axis of the interface plate. Adjust the flats and dichroic to bring the WFS axis onto the interface plate axis. 
    \item Center the lenses to the alignment telescope's axis via the crosshairs attached to the lens assembly. Correct the tilt by centering the reflection of the interface plate in L1 on the crosshairs.
    \item Repeat the previous step with the static secondary, using the central screw\footnote{This is used to mount the IR plug (a small flat mirror) onto the center of the secondary.} and reflection in the secondary mirror instead. At this point the system is in a rough alignment where it is possible to view the return beam near the focal plane on the WFS breadboard. This is useful because the alignment is very sensitive to the tilt of the secondary mirror and the return beam is only 15\% as bright as the outgoing beam. As a result, it can be challenging to find the out-of-focus return beam unless the system is well-aligned.
    \item Following the above steps, adjust the tilts of the lens assembly and the static secondary to get light into the wavefront sensor.
\end{enumerate}

The alignment telescope was useful to verify that there were no large tilts between the lens assembly and the static secondary via the reflections described in items 5 and 6 of the above list. The lens assembly and ASM mount were designed to be built to tolerance so we could not make small, repeatable adjustments to their x- and y- tilts. Although building the mounts to tolerance should hypothetically ease the alignment process, in practice we found that the alignment was extremely sensitive to the position of the ASM. Future implementations might consider adding fine tilt adjustment mechanisms into the lens and ASM assemblies, which might ease the alignment process.

\section{Static performance of IRTF-ASM-1 in the lab} \label{sec:lab_tests}

\begin{figure}[H]
    \centering
    \includegraphics[width=\linewidth]{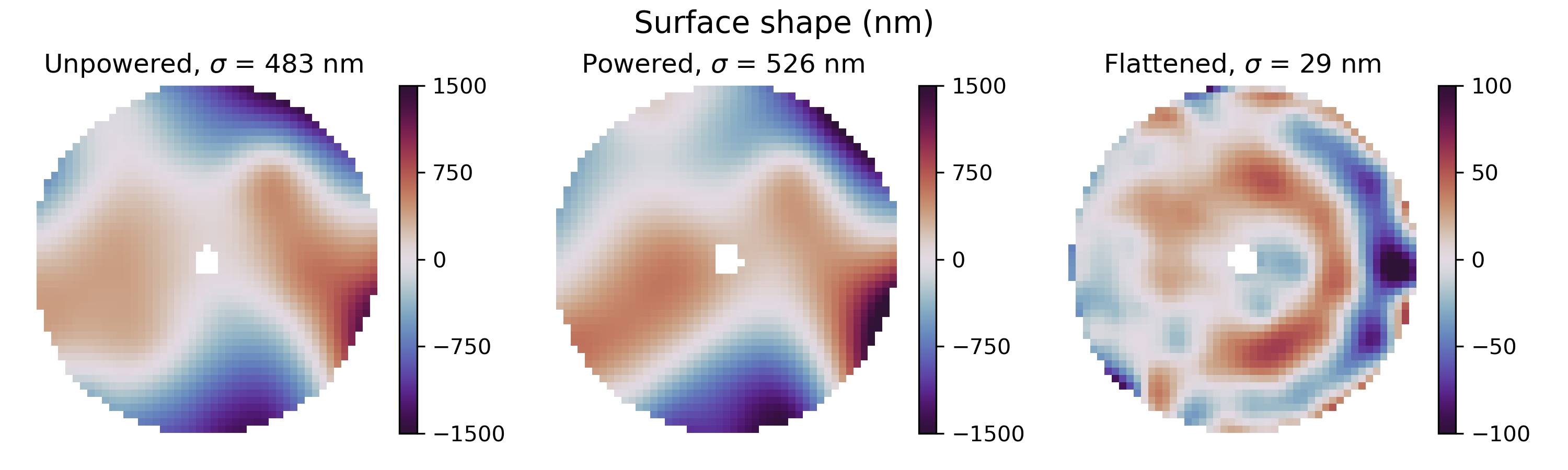}
    \caption{Static surface shapes of IRTF-ASM-1. \textit{Left:} The unpowered shape of the ASM on the test bench. \textit{Middle:} The powered , zero-command shape of the ASM, which is quite similar to the unpowered shape. \textit{Right:} The flattened, closed-loop shape of the ASM. Error terms are noted in surface figure error. }
    \label{fig:asm_surfaces}
\end{figure}

Following a pre-shipment review, IRTF-ASM-1 was hand-carried from Delft for lab testing at the IfA-Hilo facility. The ASM arrived in Hilo late in the evening on February 15th. Upon unpacking it was connected to a 64-channel set of analog electronics that provides +/- 300 mA current output to each of its outputs. Commands from our real-time computer are sent to these electronics via a 64-channel analog output PCIe board.  Six capacitive sensors, built into the ASM, were used to verify basic functionality of all 36 actuators. On February 22nd, we swapped the static secondary mirror for the ASM on the test bench. Owing to the good unpowered shape of IRTF-ASM-1 (Fig.~\ref{fig:asm_surfaces}) and experience realigning the system, we were able to recover the alignment within an hour. That evening, we took a measured interaction matrix between the wavefront sensor and ASM and closed the loop with IRTF-ASM-1 for the first time in the lab.

\subsection{Flattening the ASM}

The unpowered, powered, and flattened shapes of the ASM are shown in Fig.~\ref{fig:asm_surfaces}. In its unpowered state, the $12\times12$ Shack-Hartmann wavefront spots are easily distinguished given the relatively large 5.6-arcsecond, 10-pixel field of view of each subaperture. Some spots were outside of the capture range of the subapertures so we manually flattened the ASM before closing the loop. Note that the breadboard is also equipped with a $47\times47$ Shack-Hartmann WFS (SHWFS) produced by Imagine Optic (referred to as the "HASO WFS"), which is separate from the $12\times12$ SHWFS used to control the ASM. The HASO WFS is used to monitor the high-order shape of the ASM. The lab data that we show in this section were recorded using the HASO WFS. The flattened shape was measured by closing the loop and injecting slope offsets to remove any non-common path aberrations between the low-order WFS and the HASO WFS.

All three measurements in Fig.~\ref{fig:asm_surfaces} were performed with the ASM edge-up rather than facing downward as it does on the telescope. To first order, putting the ASM on the telescope did not have a drastic effect on the surface shape created by the flat commands. The flat commands generated in the lab gave us a reasonable image of the star in the facility guider, i.e., the image appeared to be a point source and was roughly consistent with the usual image quality of IRTF. The flat commands also provided a clear set of $12\times12$ SHWFS spots with which to close the AO loop. We plan to further investigate the effects of inclination on the ASM.

\subsection{Influence functions}

\begin{figure}[H]
    \centering
    \includegraphics[scale=0.2]{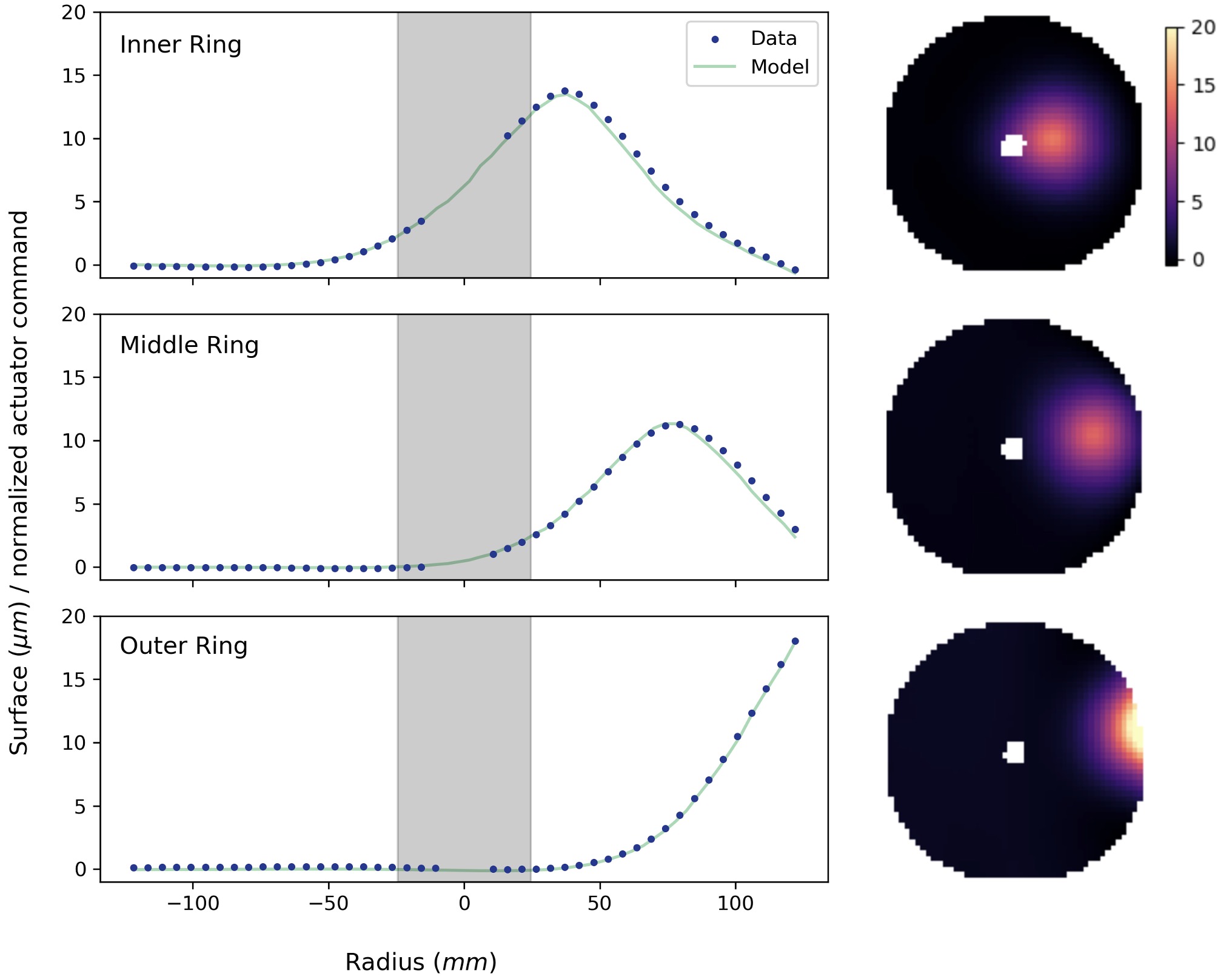}
    \caption{Measured and modeled influence functions for three actuators. The color bar on the right has the same units as the plots on the left. The shaded region indicates the central obscuration. The data points on the left are a horizontal slice of the measured influence functions on the right. Theoretical influence functions from an FEA model are also shown, with piston subtracted out. The shape of the measured influence functions are very similar to the model.}
    \label{fig:infl_func}
\end{figure}

Figure~\ref{fig:infl_func} shows differential measurements of the influence functions. We expect a different response for actuators at each of the three rings of the ASM, with actuators on the outer ring having about 1.5 times the response of actuators on the inner ring. The measured shape of the influence functions closely match simulations.

The actuator stroke is large at $\pm$\qty{15}{\micro\meter} surface deformation across the surface of the mirror. The interactuator stroke is $\pm$\qty{5}{\micro\meter} or greater and depends on the location of the actuator.\cite{2024SPIE_arjo} Using the full range of the actuators displaces the spots very far off from their expected subaperture locations. As a result, we usually try limit the actuator commands to minimize stress on the facesheet. Although we have not thoroughly stress tested IRTF-ASM-1, it was robust to human errors that exercised the large stroke of the actuators both in the lab and on-sky.

\begin{figure}
    \centering
    \includegraphics[scale=0.55]{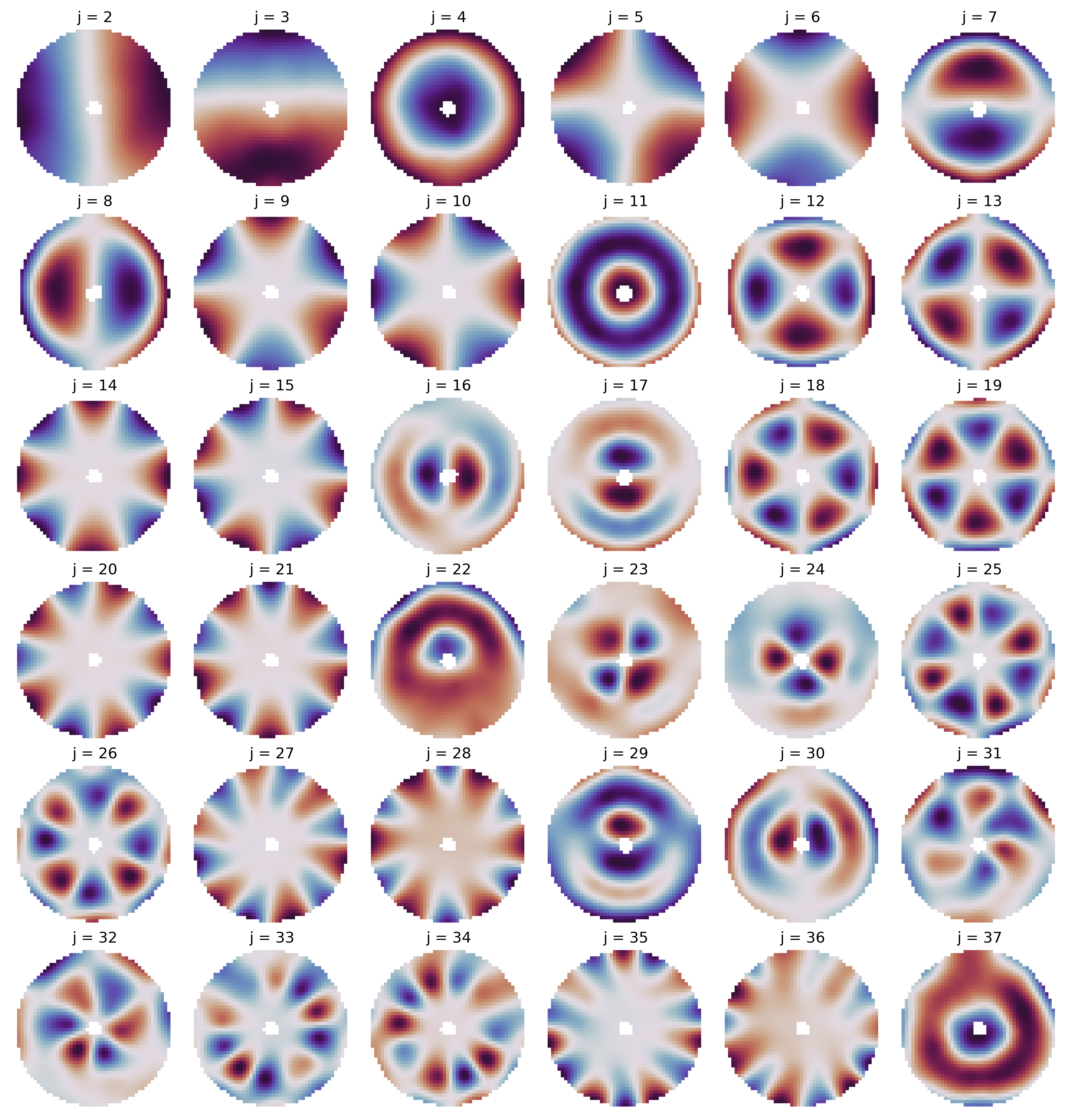}
    \caption{Noll Zernike polynomials projected on the surface of the ASM.}
    \label{fig:zernike_modes}
\end{figure}

\begin{figure}
    \centering
    \includegraphics[scale=0.55]{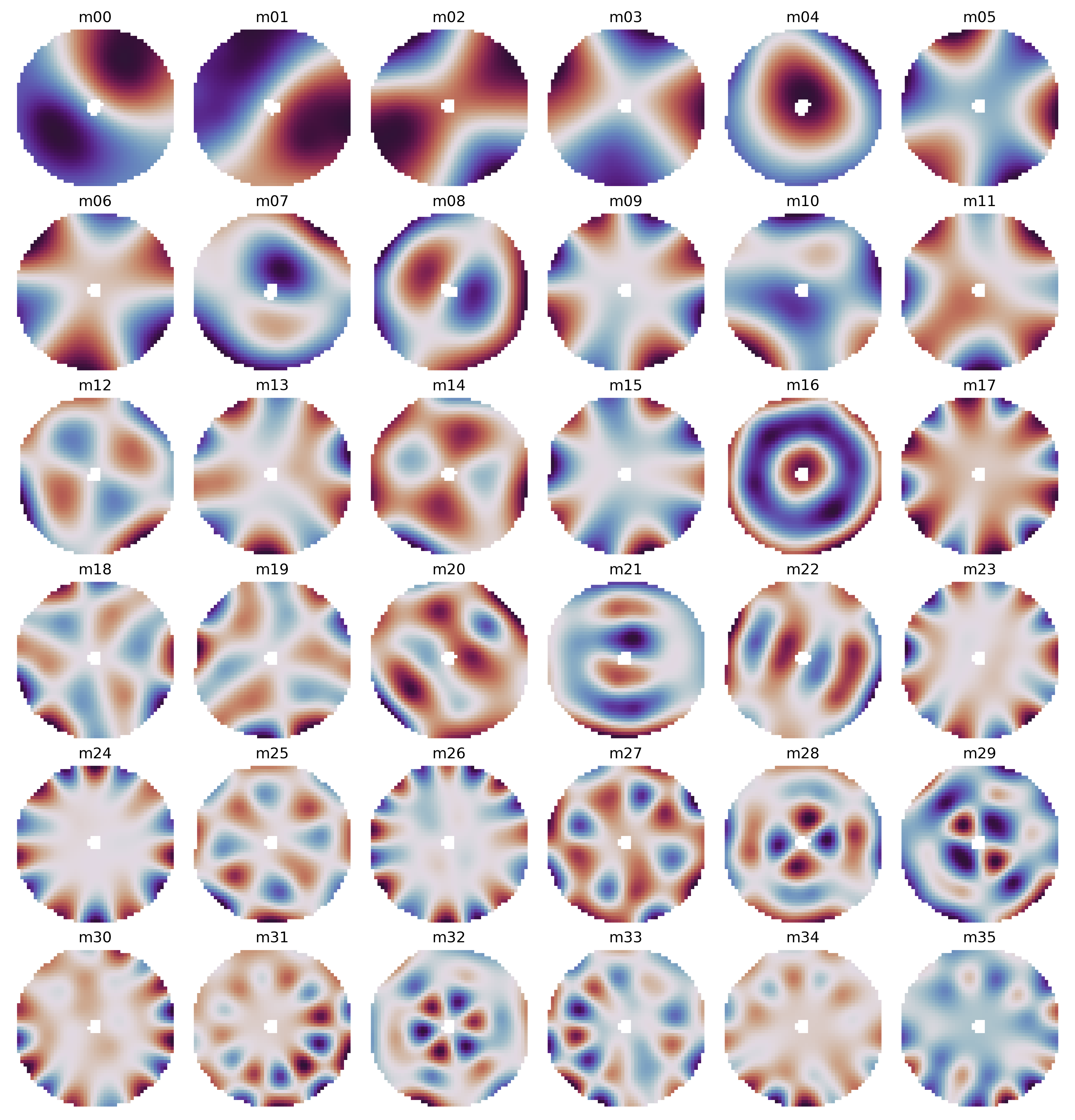}
    \caption{Mirror modes projected on the surface of the ASM.}
    \label{fig:mirror_modes}
\end{figure}

\begin{figure}
    \centering
    \includegraphics[width=\linewidth]{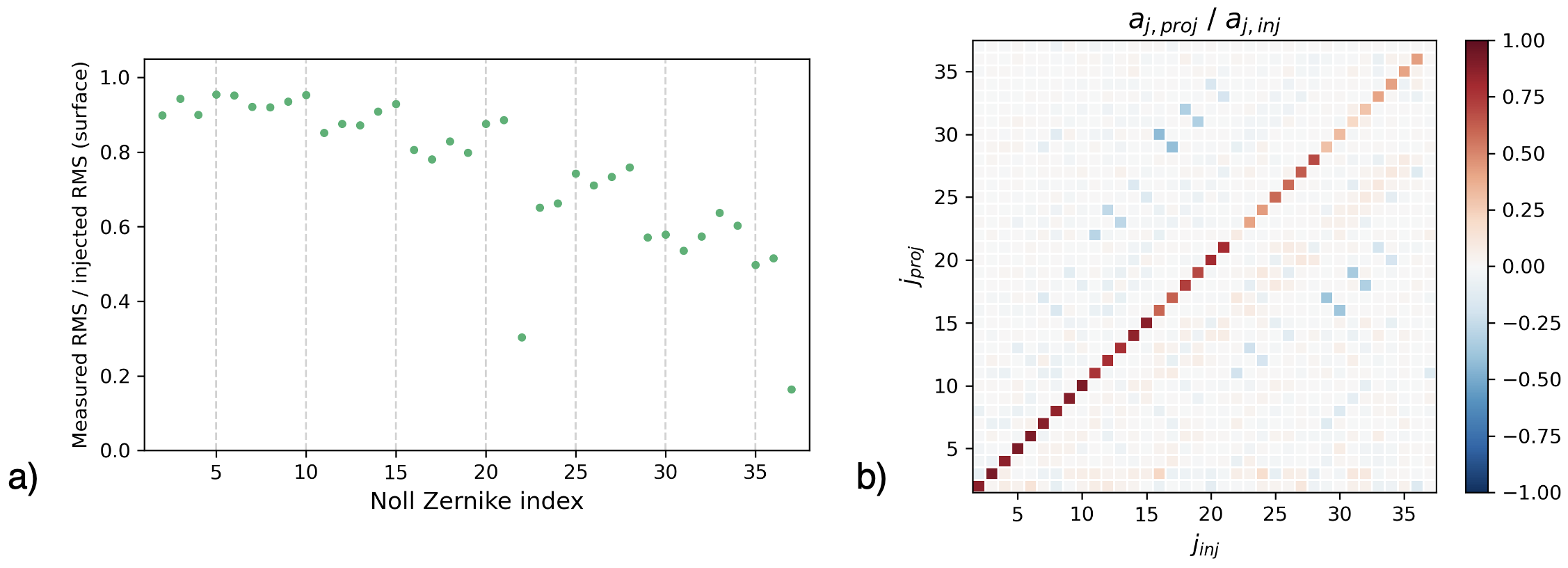}
    \caption{Input vs. output Zernike modes. a) Ratio of input to output RMS wavefront error for each Zernike mode. As expected, IRTF-ASM-1 generally struggles with higher order Zernike modes.  The actual RMS in the output wavefront clearly shows drop offs for high-order spherical terms that cannot be created with three rings of actuators. b) The output surface shape re-projected onto the Zernike basis. Each column gives the set of Zernike coefficients that are produced when we try to inject a single Zernike mode. There appears to be some covariance between the high order Zernike modes. Compare these results to the mirror modes in Fig.~\ref{fig:in_out_mirror}.}
    \label{fig:in_out_zernike}
\end{figure}

\begin{figure}
    \centering
    \includegraphics[width=\linewidth]{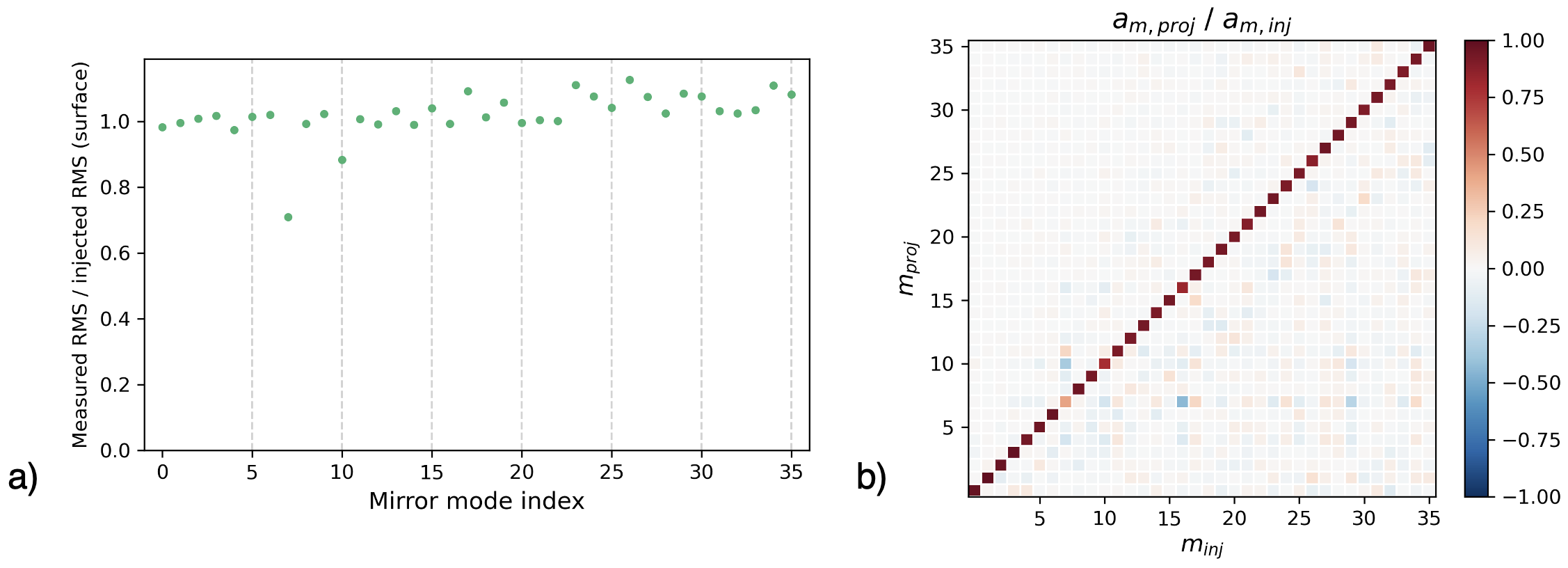}
    \caption{Input vs. output mirror modes. a) Ratio of input to output RMS wavefront error for each mirror mode. The amplitude is preserved across the majority of the mirror modes. b) The output surface shape re-projected onto the mirror mode basis. Each column gives the set of mirror mode coefficients that are produced when we try to inject a single mirror mode. With the exception of mode 7, there is less cross-talk between the mirror modes compared to the Zernike modes shown in Fig.~\ref{fig:in_out_zernike}.}
    \label{fig:in_out_mirror}
\end{figure}

\subsection{Zernike and mirror modes}

Figure~\ref{fig:zernike_modes} shows Zernike modes projected onto the ASM, which were created by projecting the influence functions of the actuators onto a basis set of 36 Zernike polynomials (not including piston). As expected, IRTF-ASM-1 struggles with high-order spherical terms because it is limited to three rings of actuators. This is apparent in Noll Zernike modes $j=22$ and $j=37$ shown in Fig.~\ref{fig:zernike_modes}. Figure~\ref{fig:in_out_zernike} also visualizes how the measured amplitude falls off with high-order modes, which we also expect. The shape of each Zernike mode is mostly preserved.

We also generated a set of mirror modes by computing the eigenmodes of the surface of the ASM from the influence functions (Fig.~\ref{fig:mirror_modes}). Using eigenmodes of the mirror is advantageous because it minimizes the number of surface shapes that are difficult for the ASM to create, as visualized in Fig.~\ref{fig:in_out_mirror}. The display of the mirror mode coefficients in Fig.~\ref{fig:in_out_mirror} are normalized in a similar manner as the corresponding plot of the Noll Zernike polynomials, such that a coefficient of 1 corresponds to \qty{1}{\micro\meter} RMS of that mirror mode. Injecting measured mirror modes is ``cleaner" than Zernike modes and does not suffer from a decrease in output amplitude with higher orders. These mirror modes were used to generate a modal control matrix that drove the ASM on-sky.

\section{On-sky performance} \label{sec:sky}
\begin{figure}[H]
    \centering
    \includegraphics[scale=0.2]{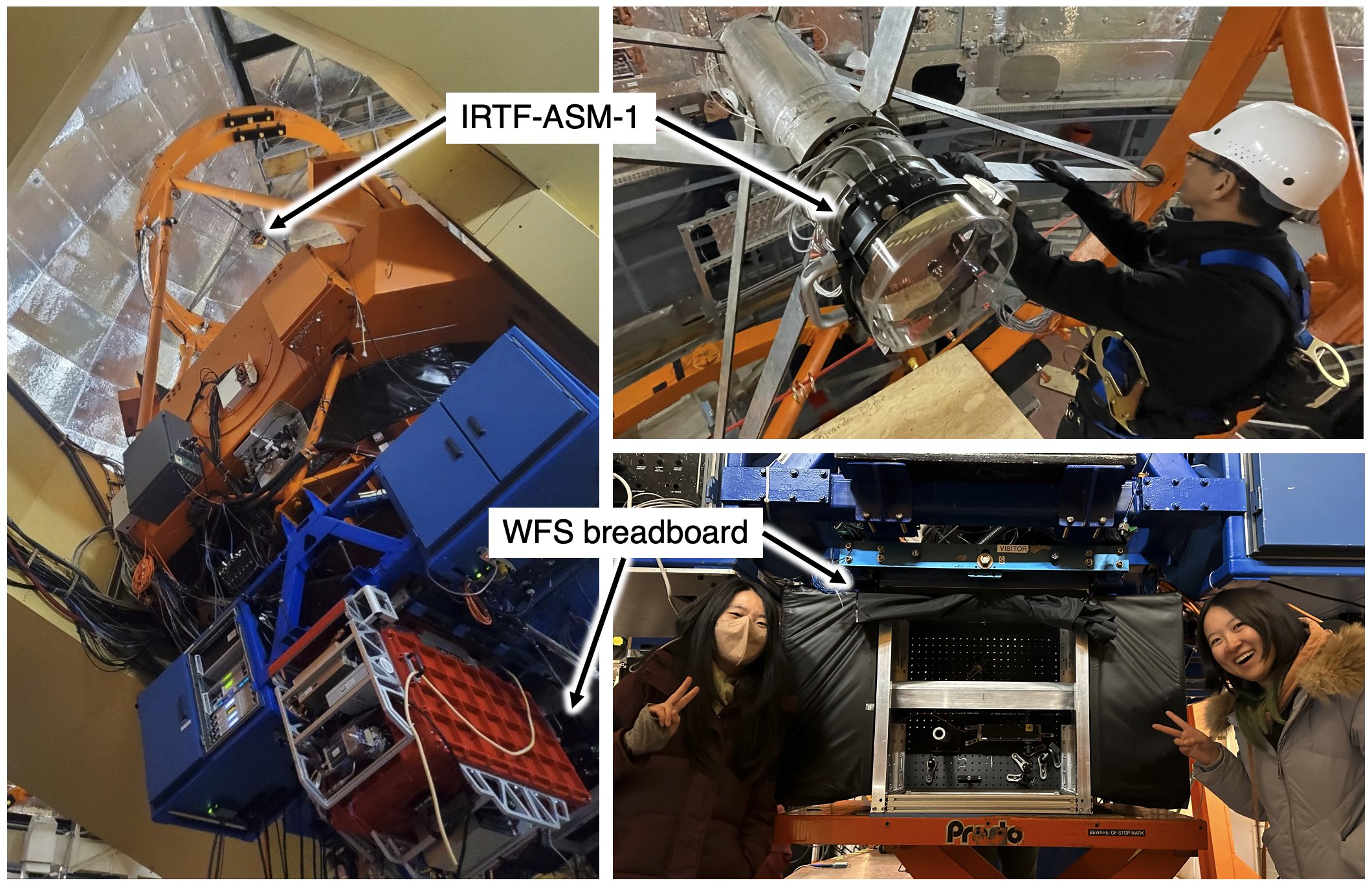}
    \caption{IRTF-ASM-1 and the wavefront sensor package on IRTF just before first light.}
    \label{fig:on_irtf}
\end{figure}

We brought the WFS breadboard and the ASM up to IRTF during the week of April 15, which was one week before the first light (Fig.~\ref{fig:on_irtf}). There were three observing runs scheduled throughout April and May on the following dates:

\begin{itemize}
    \item Run 1: April 23 - April 25, 3 nights. First light and successful closed-loop.
    \item Run 2: May 13 - May 15, 3 nights. Unable to observe due to bad weather.
    \item Run 3: May 28 - May 29, 2 nights. Better NCPA tuning, testing different control matrices and state-space controllers.
\end{itemize}

The procedure for handling the ASM at the telescope was modeled after existing static secondary mirror handling procedures and did not require any additional measures specific to the ASM other than the cabling.  The ASM was handled by the IRTF day crew and one of the engineers from the ASM project.

\begin{figure}
    \centering
    \includegraphics[width=\linewidth]{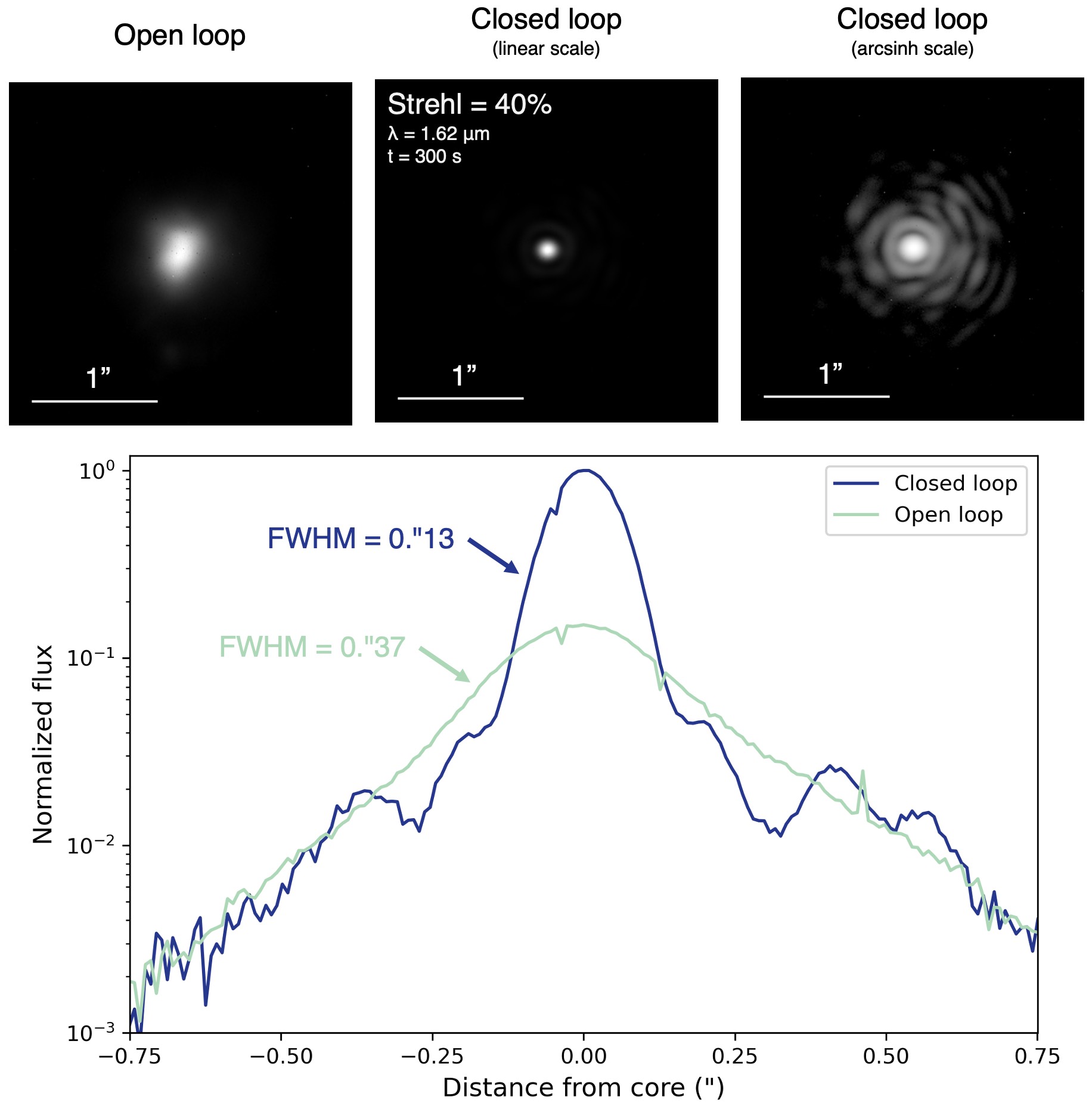}
    \caption{H-band ($\lambda=$\qty{1.62}{\micro\meter}) open- and closed-loop images of Hokule'a (alf Boo) taken on UT April 26th, 2024. The images are averaged over \qty{300}{\second} of exposure time (10 frames, \qty{30}{\second} each) and the ground wind speed was roughly 2 m/s. The bottom panel shows a comparison between the open- and closed-loop PSFs, where the plotted data are central slices of the images along the x-axis. The open-loop FWHM is 0."37 at \qty{1.62}{\micro\meter}, which is equivalent to 0."47 at \qty{0.5}{\micro\meter}. This is a typical night of good seeing on Maunakea. The open-loop FWHM was computed by fitting a gaussian.}
    \label{fig:hokulea}
\end{figure}

\begin{figure}
    \centering
    \includegraphics[width=\linewidth]{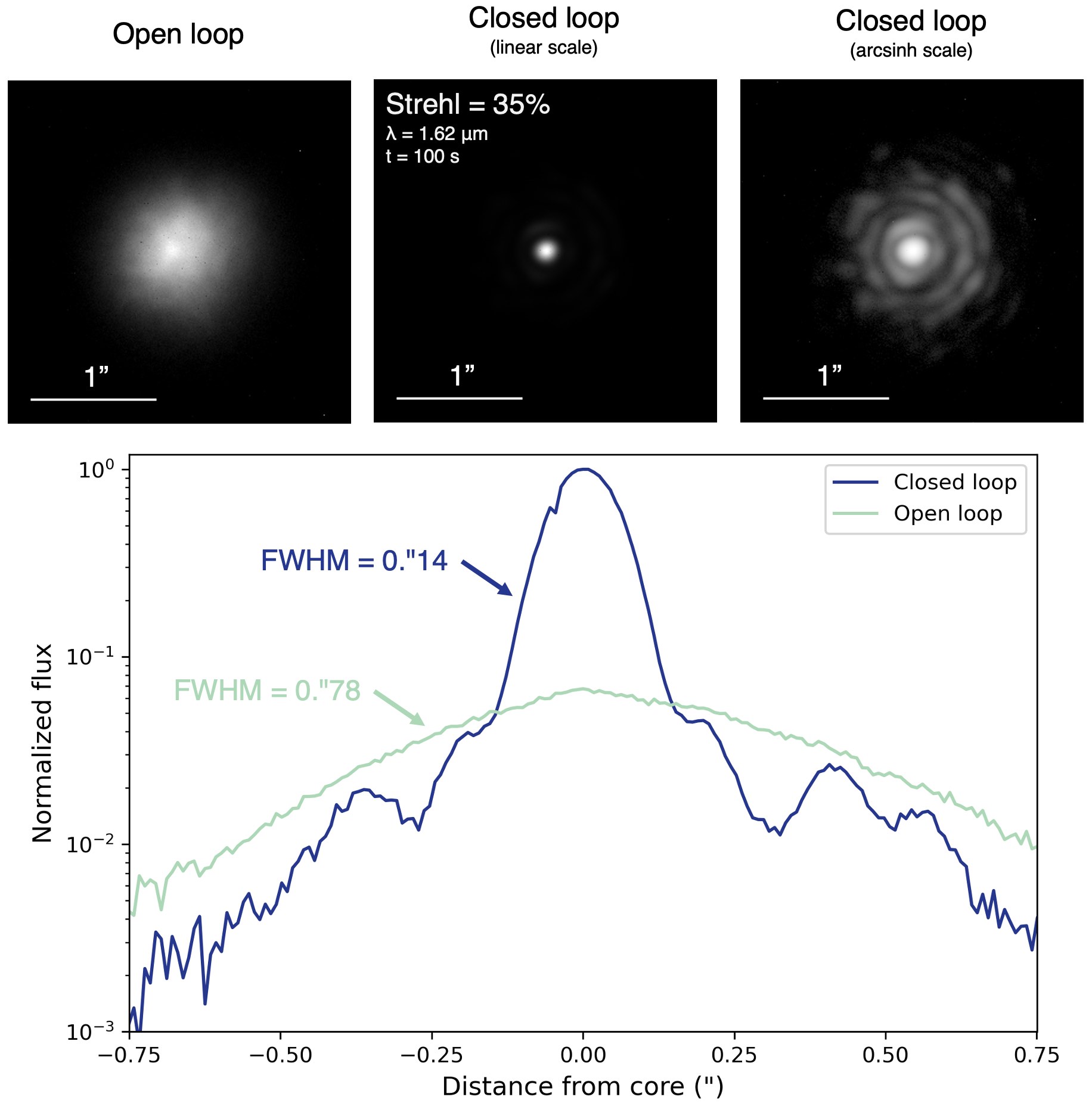}
    \caption{H-band ($\lambda=$\qty{1.62}{\micro\meter}) open- and closed-loop images of Vega taken on UT April 25th, 2024. The images are averaged over \qty{100}{\second} of exposure time (10 frames, \qty{10}{\second} each) and the ground wind speed was roughly 7 m/s. The bottom panel shows a comparison between the open- and closed-loop PSFs. The open-loop FWHM is 0."78 at \qty{1.62}{\micro\meter}, which is equivalent to 0."99 at \qty{0.5}{\micro\meter}.}
    \label{fig:vega}
\end{figure}

We closed the loop with IRTF-ASM-1 a few hours into the first night of observing. During the first run, we tested multiple methods of controlling the ASM including zonal and modal control matrices with a PID control loop as well as lead filter controller that was implemented via a state-space controller \cite{2024SPIE_suz}. Figure~\ref{fig:hokulea} shows images of Hokule'a (also known as alf Boo or Arcturus) taken in good seeing. The measured Strehl ratio is 40\% over \qty{5}{\minute} of exposure time with a wind speed of about 2 m/s measured by ultrasonic anemometers at IRTF. We measured a closed-loop full width at half maximum (FWHM) of 0."13 and an open-loop FWHM of 0."37 in H-band. This corresponds to roughly half-arcsecond seeing in visible light. The ASM performed similarly well under worse seeing conditions (1" seeing and 7 m/s winds), shown in the images of Vega in Fig.~\ref{fig:vega}. Note that the open loop images are not necessarily a fair comparison to the typical static, seeing-limited image quality of IRTF. The data were taken by first averaging over about \qty{10}{\second} of closed loop commands, and then by setting the actuator commands to the average commands. This means that the open loop images do not include static or slowly varying aberrations that are typically part of the telescope's image quality.

Interaction matrices were obtained on-sky by a method that will be described in a subsequent paper, CACOFONI (Complex Amplitude Calibration Organised in Frequency for On-sky Nimble Interaction matrix). Briefly, sinusoidal modulations are introduced at different frequencies on every actuator (or mode) either in open loop or closed loop, and the the cosine amplitudes of the centroids, measured simultaneously, are demodulated by the amplitude of the modulation, providing each column of the interaction matrix in a temporal frequency comb. The frequency step between two adjacent modulation frequencies is determined by the length of the modulation sample; we generally use 27,000 samples and we achieve a $\Delta f$ at which the peaks can easily be separated of \qty{0.1}{\hertz}. With 36 simultaneous actuators or modes, we generate a frequency comb between 5 and \qty{8.6}{\hertz}, within a frequency range where the ASM open loop response is most linear. We tested both zonal and modal modulations, in open and closed loop. We also used these measured matrices to generate synthetic interaction matrices, based on a model of the wavefront sensor and the measured influence functions, which was adjusted to fit to the measured on-sky matrices by varying geometrical parameters ($\Delta x$, $\theta$, magnification) by an amoeba algorithm. No significant difference in closed loop performance was noted with these different control matrices.

The combined emissivity of IRTF-ASM-1 and the primary mirror was 6\%, which is within normal emissivity measurements at IRTF (between 5\% and 6\% with the static secondary). We took this measurement with the pupil viewing mode of the facility near-infrared spectrograph SpeX. We pointed to the sky and compared the L-band flux from ASM to the flux from a reference material with an emissivity of 100\%. Continuously moving the actuators while performing the measurement did not change the emissivity.

\section{Conclusions \& future work}

IRTF-ASM-1 has successfully demonstrated the TNO hybrid variable reluctance actuator on-sky.  It achieved stable diffraction-limited performance with bright-star long-exposure Strehl ratios of 35-40\% in H-band under 0.5-1 arcsecond seeing conditions over a number of engineering nights.  Despite the compressed timeline of this project, the ASM met all of its performance requirements and proved to be highly robust under operation at an astronomical facility. Between shipping the ASM from the Netherlands to Hawai'i, transporting it to the summit, and performing multiple top end exchanges over the course of a month, we did not experience any degradation of performance or unexpected behavior from the ASM.

In the coming months, we plan to regularly exercise IRTF-ASM-1 at the summit to evaluate its long-term robustness in a telescope dome environment. IRTF will also commission FELIX, a new facility guider and combined Shack-Hartmann wavefront sensor, which is scheduled to go on-sky this fall. FELIX was originally designed to monitor low-order, slowly varying telescope aberrations but could potentially evaluate high-order aberrations via extrafocal images \cite{tokovinin2006donut}. Preliminary tests of correcting static high-order and fast low-order aberrations (binned down to $3\times 3$ subapertures) with the ASM show promising results, providing a Strehl of around 20\% at H-band over \qty{300}{\second} of exposure time in 0."5 seeing. Although IRTF's instruments are not designed to sample a diffraction-limited PSF, wavefront correction from the ASM will improve the throughput/sensitivity of facility spectrographs.

\acknowledgments 
 
This work and two of its authors (Lee and Zhang) are supported by a National Science Foundation Advanced Technology and Instrumentation award (NSF-1910552). The Infrared Telescope Facility is operated by the University of Hawaii under contract 80HQTR19D0030 with the National Aeronautics and Space Administration. We would like to thank TNO for collaborating on the ASM project. We would also like to thank the IRTF staff smooth and successful operations at the telescope. We also thank our collaborators at UCSC for slumping IRTF-ASM-1's facesheet as well as Huygens Optics for polishing the mirror. We wish to recognize and acknowledge the very significant cultural role and reverence that the summit of Maunakea has always had within the Native Hawaiian community. We are most fortunate to have the opportunity to conduct observations from this mountain. 

\bibliography{report} 

\begin{thebibliography}{10}

\bibitem{2020SPIE11448E..5VB}
{Bowens-Rubin}, R., {Hinz}, P., {Jonker}, W., {Kuiper}, S., {Laguna}, C., and {Maniscalco}, M., ``{Performance of large-format deformable mirrors constructed with TNO variable reluctance actuators},'' in [{\em Adaptive Optics Systems VII}{\nolinebreak\hspace{0.1em}]},  {Schreiber}, L., {Schmidt}, D., and {Vernet}, E., eds., {\em Society of Photo-Optical Instrumentation Engineers (SPIE) Conference Series} {\bf 11448},  114485V (Dec. 2020).

\bibitem{2021SPIE11823E..1RB}
{Bowens-Rubin}, R., {Hinz}, P., {Kuiper}, S., and {Dillon}, D., ``{Performance of large-format deformable mirrors constructed with hybrid variable reluctance actuators II: initial lab results from FLASH},'' in [{\em Techniques and Instrumentation for Detection of Exoplanets X}{\nolinebreak\hspace{0.1em}]},  {Shaklan}, S.~B. and {Ruane}, G.~J., eds., {\em Society of Photo-Optical Instrumentation Engineers (SPIE) Conference Series} {\bf 11823},  118231R (Sept. 2021).

\bibitem{bowens2023performance}
Bowens-Rubin, R., Marsh, A., Hinz, P., Laguna, C., Bos, A., Kuiper, S., Baeten, M., and Dillon, D., ``Performance of large-format deformable mirrors constructed with hybrid variable reluctance actuators iii: laboratory measurements of dynamic behavior,'' in [{\em Adaptive Optics for Extremely Large Telescopes 7th Edition}{\nolinebreak\hspace{0.1em}]},  (2023).

\bibitem{kuiper2019adaptive}
Kuiper, S., Jonker, W., Maniscalco, M., Priem, H., Coolen, C., Chun, M., Baranec, C., Lu, J., and Lai, O., ``Adaptive secondary mirror development for the uh-88 telescope,'' in [{\em AO4ELT6}{\nolinebreak\hspace{0.1em}]},  (2019).

\bibitem{2020SPIE11448E..1EC}
{Chun}, M., {Baranec}, C., {Lai}, O., {Lu}, J.~R., {Zhang}, R., {Kuiper}, S., {Jonker}, W., and {Maniscalco}, M., ``{A new adaptive secondary mirror for astronomy on the University of Hawaii 2.2-meter telescope},'' in [{\em Adaptive Optics Systems VII}{\nolinebreak\hspace{0.1em}]},  {Schreiber}, L., {Schmidt}, D., and {Vernet}, E., eds., {\em Society of Photo-Optical Instrumentation Engineers (SPIE) Conference Series} {\bf 11448},  114481E (Dec. 2020).

\bibitem{2022SPIE12185E..7UC}
{Chun}, M.~R., {Ryan}, A., {Zhang}, R., {Kuiper}, S., {Ackaert}, G., {Baranec}, C., {Baeten}, M.~J.~J., {Bos}, A., {Bowens-Rubin}, R., {Dekker}, B., {Dungee}, R., {Gupta}, T., {Hinz}, P., {Jonker}, W., {Kamphues}, F., {Lai}, O., {Lu}, J., {Maniscalco}, M., {Monna}, B., {Nair}, M., {Nijenhuis}, J., {Priem}, H., and {Vogel}, P.-A., ``{Progress on the University of Hawaii 2.2-meter adaptive secondary mirror},'' in [{\em Adaptive Optics Systems VIII}{\nolinebreak\hspace{0.1em}]},  {Schreiber}, L., {Schmidt}, D., and {Vernet}, E., eds., {\em Society of Photo-Optical Instrumentation Engineers (SPIE) Conference Series} {\bf 12185},  121857U (Aug. 2022).

\bibitem{hinz2020developing}
Hinz, P.~M., Bowens-Rubin, R., Baranec, C., Bundy, K., Chun, M., Dillon, D., Holden, B., Jonker, W., Kosiarek, M., Kupke, R., et~al., ``Developing adaptive secondary mirror concepts for the apf and wm keck observatory based on hvr technology,'' in [{\em Adaptive Optics Systems VII}{\nolinebreak\hspace{0.1em}]},   {\bf 11448},  1172--1187, SPIE (2020).

\bibitem{bowens2022adaptive}
Bowens-Rubin, R., Bos, A., Hinz, P., Holden, B., and Radovan, M., ``An adaptive optics upgrade for the automated planet finder telescope using an adaptive secondary mirror,'' in [{\em Adaptive Optics Systems VIII}{\nolinebreak\hspace{0.1em}]},   {\bf 12185},  579--593, SPIE (2022).

\bibitem{2024SPIE_arjo}
{Bos}, A., {Dekker}, B., {Kuiper}, S., {Kidron}, M., {Salcedo}, E.~V., {Baeten}, M., {Vermeulen}, R., {Kuijt}, J., {Boot}, K., {van Venrooy}, B., {van Buuren}, R., {Kamphues}, F., {Jonker}, W.~A., {Maniscalco}, M., {Chun}, M.~R., {Connelley}, M.~S., {Ryan}, A., {Lee}, E., {Zhang}, R., {Lai}, O., {Vleggaar}, J. J.~M., and {Hinz}, P.~M., ``{Status and first results of the NASA IRTF Adaptive Secondary Mirror},'' in [{\em Adaptive Optics Systems IX}{\nolinebreak\hspace{0.1em}]},  {\em Society of Photo-Optical Instrumentation Engineers (SPIE) Conference Series} (2024).

\bibitem{2024SPIE_suz}
{Zhang}, R., {Baeten}, M., {Chun}, M.~R., {Lee}, E., {Connelley}, M., {Lai}, O., {Kuiper}, S., {Ryan}, A., {Bos}, A., {Bowens-Rubin}, R., and {Hinz}, P.~M., ``{Close-loop results of adaptive secondary mirrors with TNO's hybrid variable reluctance actuators},'' in [{\em Adaptive Optics Systems IX}{\nolinebreak\hspace{0.1em}]},  {\em Society of Photo-Optical Instrumentation Engineers (SPIE) Conference Series} (2024).

\bibitem{2024SPIE_phil}
{Hinz}, P.~M., {Falk}, J., {Stelter}, D.~R., {Radovan}, M.~V., and {Dillon}, D., ``{Progress on creating thin curved glass facesheets for deformable mirrors via free form slumping},'' in [{\em Adaptive Optics Systems IX}{\nolinebreak\hspace{0.1em}]},  {\em Society of Photo-Optical Instrumentation Engineers (SPIE) Conference Series} (2024).

\bibitem{2008SPIE.7015E..65S}
{Stalcup}, T., J., {Hinz}, P., {Durney}, O., {Connors}, T., and {Mopidevi}, R., ``{A test stand for the MMT Observatory adaptive secondary},'' in [{\em Adaptive Optics Systems}{\nolinebreak\hspace{0.1em}]},  {Hubin}, N., {Max}, C.~E., and {Wizinowich}, P.~L., eds., {\em Society of Photo-Optical Instrumentation Engineers (SPIE) Conference Series} {\bf 7015},  701565 (July 2008).

\bibitem{1974ESOTR...3.....W}
{Wilson}, R.~N., ``{Test methods for secondary mirrors of Cassegrain telescopes with special reference to the ESO 3.6 m telescope.},'' {\em ESO Tech. Rep}~{\bf 3} (Jan. 1974).

\bibitem{tokovinin2006donut}
Tokovinin, A. and Heathcote, S., ``Donut: measuring optical aberrations from a single extrafocal image,'' {\em Publications of the Astronomical Society of the Pacific}~{\bf 118}(846),  1165 (2006).

\end{thebibliography}
\bibliographystyle{spiebib} 

\end{document}